\documentclass[conference]{IEEEtran}

\usepackage{graphicx}
\usepackage{epsf}
\usepackage{mathtools}
\usepackage{amssymb}
\usepackage{amsmath, amsfonts}
\usepackage{latexsym}
\usepackage{multirow}
\usepackage{multicol}
\usepackage[table]{xcolor}
\usepackage{color}

\usepackage{tabularx}
\usepackage{lipsum}

\usepackage{algorithm}
\usepackage{algorithmic}

\usepackage{mathrsfs}

\usepackage{fixltx2e}

\usepackage[mathscr]{euscript}

\usepackage{commath}
\usepackage{MnSymbol}

\usepackage{subcaption}

\usepackage{mathtools} 
\DeclarePairedDelimiterX\setc[2]{[}{]}{\,#1 \;\delimsize\vert\; #2\,}
\DeclarePairedDelimiterX\parth[2]{(}{)}{\,#1 \;\delimsize\vert\; #2\,}

\PassOptionsToPackage{hyphens}{url}
\usepackage[unicode=true, bookmarks=true, bookmarksnumbered=true, bookmarksopen=true, bookmarksopenlevel=1, breaklinks=false, pdfborder={0 0 0}, pdfborderstyle={}, backref=false, colorlinks=false]{hyperref}

\usepackage{theorem}
{
\theorembodyfont{\rmfamily}

}

\usepackage{soul}

\definecolor{orange}{RGB}{255,127,0}
\definecolor{blue}{RGB}{0,0,255}
\definecolor{red}{RGB}{255,0,0}
\definecolor{green}{RGB}{50,160,50}
\definecolor{grey}{RGB}{125,120,125}
\definecolor{purple}{RGB}{125,0,125}

\newcommand{\MK}[1]{{\color{black}{#1}}}
\newcommand{\BJ}[1]{{\color{black}{#1}}}

\begin{document}
{
\title{{\fontsize{20}{2}\selectfont Can 5G Coexist with Satellite Uplink in 28 GHz Band?}}

\author
{
Bryce Jeffrey and Seungmo Kim, \textit{Senior Member}, \textit{IEEE}


\thanks{B. Jeffrey and S. Kim are with the Department of Electrical and Computer Engineering, Georgia Southern University, Statesboro, GA, USA.}
 }

\maketitle
\begin{abstract}
\MK{5G standalone (SA) rollout is right around the corner. 28 GHz band is considered as one of the main spectrum bands for the 5G SA in many countries. However, the band has already been occupied by uplink of the fixed satellite service (FSS). Due to high equivalent isotropic radiated power (EIRP) adopted by the FSS, the interference that FSS may cause into 5G is garnering research interest. This research aims to establish an analytical framework that quantifies the FSS-to-5G interference.}
\end{abstract}


\section{Introduction}\label{sec_intro}
\subsubsection{Background}\label{sec_intro_background}
\MK{
The spectrum of 27.5-28.35 GHz (the ``28 GHz band'') \cite{47cfr30} was auctioned by the United States (U.S.) Federal Communications Commission (FCC) in 2019 for the deployment of 5G, whose aggregated bidding amount surpassed \$7 million \cite{fcc_auction}.

However, the 28 GHz band had already been occupied by the uplink of fixed satellite service (FSS), which facilitates the use of satellite communication systems that provide fixed links between ground-based earth stations (ESs) and satellites. Due to the longevity that it has to serve, an ES usually adopts a very high equivalent isotropic radiated power (EIRP) \cite{mdpi23}. As such, an interference problem from FSS ES to 5G arises.
}

\subsubsection{Related Work}\label{sec_intro_related}
\MK{
Inter-technology interference has been thoroughly studied in the literature of wireless communications. We found the following prior studies particularly relevent to this work: coexistence between Wi-Fi and military rader \cite{dietrich_lett17}, and between Long Term Evolution (LTE) and pulsed radar \cite{psun_icnc15}-\cite{dietrich_wcnc16} in spectrum of 3.550-3.700 GHz (also known as the ``3.5 GHz band); coexistence between vehicle-to-everything (V2X) and public safety communications in spectrum of 4.940-4.990 GHz (also known as the ``4.9 GHz Public Safety Band'') \cite{faizan_iceic22}; coexistence between WiGig and New Radio Unlicensed (NR-U) in spectrum of 57-71 GHz (also known as the ``60 GHz band'') \cite{kabir_hst19}\cite{verboom_vtc21s}; coexistence between V2X communications and Wi-Fi in spectrum of 5.850-5.925 GHz (also known as the ``5.9 GHz band'') \cite{dietrich_gc18}\cite{dessalgn_milcom19}.

Among them, albeit not very large, there is a body of literature on interference between 5G and FSS. However, the vast majority of the prior art is focused on the C-Band spectrum (i.e., 4-8 GHz). Interference from 5G to be deployed across the spectrum of 3.3–3.6 GHz in many countries was studied \cite{mdpi23}\cite{access22}\cite{tvt22}.

Only few prior studies such as \cite{jsac17} have discussed the FSS-to-5G interference in 28 GHz band. It only discussed FSS-to-5G base station (BS) interference for the reason that interference observed at the BSs is expected to be the major bottleneck for 5G system deployments. The rationale was that UEs likely to have smaller antenna gains and experience much higher propagation losses from the ES transmitters than APs. Thus, addressing this limitation, this paper sheds light on the interference from FSS ES to 5G user equipment (UE).

Furthermore, most of the prior studies adopt the interference-to-noise ratio (INR) as the metric to measure the level of interference. As an effort to promote clarity, this paper adopts signal-to-interference-plus-noise ratio (SINR) as the metric.
}

\subsubsection{Contributions}
\MK{Addressing the limitations of the aforementioned prior work, this research highlights the following contributions:
\begin{itemize}
\item Building a simulator aiming to measure the FSS-to-5G interference
\item Drawing suggestions on reasonable separation distances between 5G and FSS
\end{itemize}
}

\section{System Model}\label{sec_model}
\BJ{
The analysis stems from a multifaceted selection of criteria to define SINR observed at a 5G receiver (Rx). Specifically for these results, the criteria involved can be divided into two input sets to define the UE Rx and the interfering ES transmitter (Tx) within the simulator. Based on these inputs, an illustration could be made to represent the resultant Rx SINR against its distance from the interfering Tx. For all data sets, the terrestrial propagation between the FSS ES Tx(s) and the 5G UE Rx assumed a free-space path loss (FSPL) model in line-of-sight (LoS) conditions.
}

\subsection{5G UE: The Victim Rx}\label{sec_receiver_model}
\BJ{
The main inputs relating to the Rx are defined as the synchronization signal reference signal received power (SS-RSRP) and the noise factor. Here, an operating bandwidth of 1 GHz was assumed with a standard thermal temperature of 290 K to compute the noise. The RSRP was based on an average minimum power within the accepted range of what is considered \MK{an acceptable} signal strength \cite{mcta22}\cite{scte19}, resulting in an RSRP \MK{of -80 dBm}.
}

\subsection{FSS ES: The Interfering Tx}\label{sec_transmitter_model}
\BJ{
The simulator is designed to incorporate different levels of variation in regard to the Tx to obtain a more realistic result. One notable variable is the amount of Txs introduced to observe the effect of a combined interference. Namely, increasing the quantity from a single Tx to five and ten with respect to the same distance. Note that this scenario assumes all Tx are equidistant from each other and the Rx while adhering to the same FSPL and LoS conditions, therefore providing the same level of interference for each additional Tx added.

Another important distinction is the difference in the effect of interference regarding the angle and direction of propagation from the Tx. For these simulations, two data sets were incorporated based on the assumption that the interference originated from either the \MK{main lobe or a sidelobe experiencing maximum attenuation of -30 dB}. From these parameters, an estimated EIRP could then be defined for \MK{Classes 1, 2, and 3 \cite{jsac17} given by 42.2, 54.1, and 78 dBm/GHz for the mainlobe and 12.2, 24.1, and 48 dBm/GHz for the sidelobe, respectively.}
}


\section{Experimental Results}\label{sec_results}

\BJ{
Fig. \ref{fig_class_1} depicts the effect that the Class-1 ES imposes on the 5G UE's perceived SINR under the conditions mentioned in Section \ref{sec_model}. The notable attributes from this class of ES show a negligible interference level when observing the sidelobe power, regardless of the amount of Txs present, while the mainlobe introduces a more pronounced dip in performance. This effect can be seen up to 1500 m for a single Tx and upwards of 2500 m for additional Txs.

Fig. \ref{fig_class_2} depicts the results of a similar simulation with instead the utilization of the Class-2 ES. When considering the sidelobe interference, this class follows a similar trend to the Class-1 ES, with an insignificant effect past 500 m. The SINR continues to suffer, however, when examining the interference from the mainlobe, with an individual Tx producing a noticeable drop past 2500 m.

Fig. \ref{fig_class_3} depicts the most prominent levels of interference with the implementation of the Class-3 ES. Under these high-power conditions, the effect from the sidelobe of a single Tx is noticeable up to 2500 m, with additional Txs producing an even greater effect. Any interference from the mainlobe could render a received signal to the Rx UE imperceptible.

It is worth noting that for most instances, a UE would not be within the main lobe or beam of an FSS ES and the sidelobe effects would be more common. These mainlobe simulations are to consider a worst-case scenario to better suggest reasonable separation distances for each class of ES.
}

\MK{
Assuming that SINR needs to be maintained above 0 dB, one can infer the minimum separation distance between the interfering FSS ES and the victim 5G UE. That is, from Figs. \ref{fig_class_1} through \ref{fig_class_3}, it can be acknowledged that Class-1 through -3 ESs must be separated from 5G UEs by at least 100, 500, and 2500 m, respectively. Notice that the separation distance can vary according to the number of interfering FSS ESs.
}

\begin{figure}[t]
    \centering
    \includegraphics[width=\linewidth]{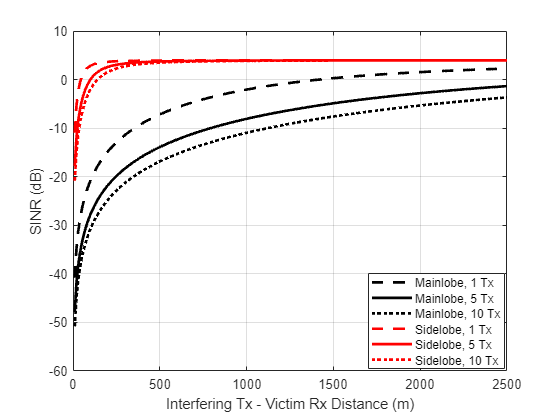}
    \caption{Class-1 ES-to-5G UE interference}
    \label{fig_class_1}
\end{figure}

\begin{figure}[t]
    \centering
    \includegraphics[width=\linewidth]{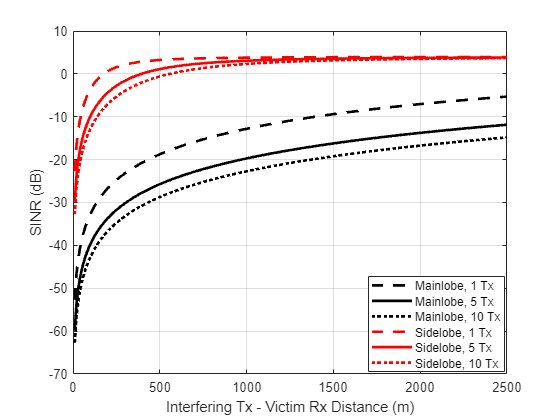}
    \caption{Class-2 ES-to-5G UE interference}
    \label{fig_class_2}
\end{figure}

\begin{figure}[t]
    \centering
    \includegraphics[width=\linewidth]{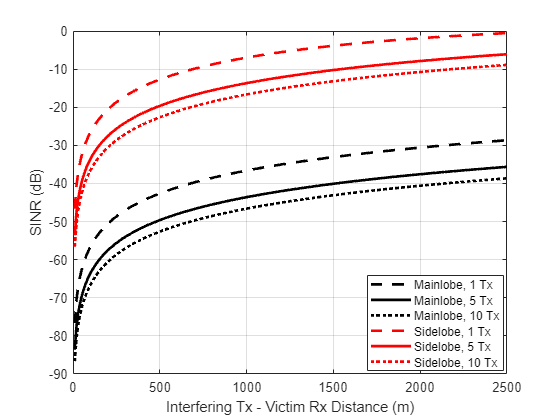}
    \caption{Class-3 ES-to-5G UE interference}
    \label{fig_class_3}
\end{figure}

\section{Concluding Remarks}\label{sec_conclusions}
\MK{This paper has presented a set of simulation results on the 5G received SINR. It identified the FSS ES and the 5G UE as the interfering Tx and the victim Rx, respectively, and quantified separation distances between FSS ES and 5G UE for Classes 1, 2, and 3. This research will be improved such that more realistic interference scenarios are taken into account.}


\end{document}